# EPICS DEPLOYMENT AT FERMILAB[*]

P. Hanlet[†], M. Gonzalez, J. Diamond, K.S. Martin
Fermi National Accelerator Laboratory, Batavia, Illinois, USA

*Abstract*

Fermilab has traditionally not been an EPICS house; as such expertise in EPICS is limited and scattered. PIP-II will be using EPICS for its control system. When in operation, it will need to interface with the existing, modernized (see ACORN) legacy control system. Treating EPICS controls at Fermilab as a green field, we have developed and deployed a software pipeline which addresses these needs and presents to developers a tested and robust software framework, including template IOCs from which new developers can quickly deploy new front ends, aka IOCs. In this presentation, motivation for this work, implementation of a continuous integration/continuous deployment pipeline, testing, template IOCs, and the deployment of user services/applications will be discussed. This new infrastructure of IOCs and services is being developed and used in the PIP-II cryomodule teststand; our experiences and lessons learned will be also be discussed.

## INTRODUCTION

Fermi National Accelerator Laboratory, or Fermilab, in Batavia, Illinois, USA is constructing a new superconducting RF (SRF) linear accelerator (LINAC) with twice the energy of the existing LINAC and significantly higher power. This LINAC, known as PIP-II, will power the rest of the Fermilab accelerator complex, generating the world's most intense high-energy neutrino beam, as well as providing beam to other experiments and test beams, see Fig. 1.

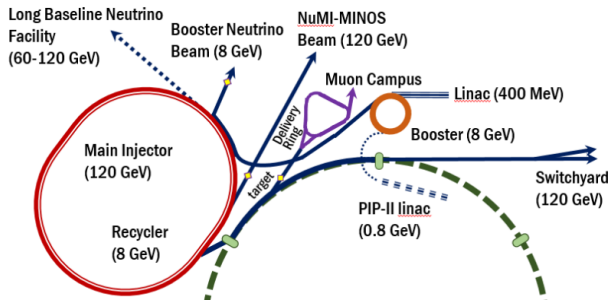

Figure 1: Fermilab's accelerator chain and experiments

The capabilities of PIP-II [1, 2] are to provide an 800 MeV proton beam of 1.2 MW using a superconducting RF LINAC, see Fig. 2. The beam is CW-compatible and customizable for a variety of user requirements, and is upgradeable to multi-MW power. The scope of PIP-II includes a beam transfer line to the existing Booster ring and accelerator complex upgrades to the Booster, Recycler, Main Injector, and conventional facilities. Space is also reserved in the LINAC for 2 additional SRF cryomodules (CMs). For more on PIP-II, see [3].

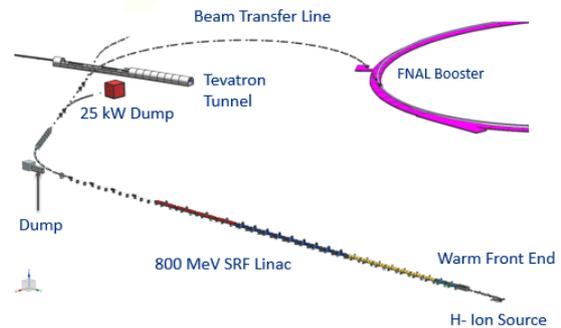

Figure 2: PIP-II scope: new LINAC and beam transfer line to Booster

## GOALS FOR DEPLOYMENT

Though EPICS [4] has been long been deployed in small pockets at Fermilab, one could say that Fermilab has never been an EPICS shop. As such, the expertise is limited. Furthermore the controls engineers who support the remainder of the accelerator complex are stretched thinly. Therefore, for a small controls team, we want to deploy EPICS in a sustainable way by requiring 1) a robust build of infrastructure, 2) automated build procedures, 3) extensive testing, and 4) minimal functionality to automate testing, deployment, and monitoring of IOCs.

We are treating EPICS deployment as a green field to simplify development for non-experts. This follows from the fact PIP-II, and likely new components from ACORN [5], will not rely on old hardware, such as VME and CAMAC. With this in mind, we use current versions of EPICS software, both for base and Support modules, with an expectation that new software versions will continue to operate for the foreseeable future. This is an ideal environment to adopt modern computing procedures, in particular employing a Continuous Integration/Continuous Deployment (CI/CD) pipeline. This allows us to build all of EPICS base and EPICS Support/Modules as "standard" Fermilab versions and deploy it on an NSF host for all to use. Furthermore, we have developed "template IOCs" which have the boiler plate functionality which we require for all IOCs on the Fermilab controls network.

In addition to standardizing code, we are exploring standardizing hardware platforms for hosting IOCs. We are using

---

Buildroot to build for raspberry pis, and two System on Module (SoM) platforms: the arm/Cyclone-V and arm/Arria-10 platforms. As of the writing of this document, we are adding Yacto for building Xilinx-based SoM devices. Both Buildroot and Yacto will produce kernels and root file systems for each platform; these are required for network booting embedded nodes. Additionally, these tools generate Toolchains, or SDKs, for cross compiling the entire EPICS code base.

At the time of writing this article, we are using EPICS base-R7.0.7.

*PVXS Protocol*

The newest implementation of the pvAccess (PVA) api, called 'PVXS", has been selected in lieu of the traditional Channel Access (CA) so as to take advantage of its more efficient network protocol, which handles higher bandwidth data and its support of structured data and "normative types", or specific structure instances of pvData structures. Another advantage of PVXS is its native support of multicast, which is essential on Fermilab's extensive controls network which has numerous VLANs across which broadcast won't work. To fulfill the impending government requirement that networks use IPv6, PVXS is also the protocol of choice. Lastly, with the recognition that network security, and future restrictions looming regarding network security, Fermilab looks to be at the forefront of implementing and testing the newest measures presently under development [6, 7].

## TEMPLATE IOCS

Each Fermilab IOC running on the controls network will be required to provide the following functionalities: a heartbeat, IOC statistics (CPU usage, memory usage, etc.), the capability of 20 $Hz$ scan rate (the PIP-II pulse frequency), access to acnet (acnetPV wrapper), use aSub (specific Support module) record for interfacing with IOC specific custom code (C/C++ libraries; e.g. acnetPV), tcast (an interface to the clock system), and reccaster (for use interfacing with Channel Finder).

These capabilities will be used in the functional testing step of the template IOCs in the CI/CD pipeline. Furthermore, when the IOCs are deployed to production, these tests will be used to monitor IOC health.

Template IOCs are kept in the iocTops directory. There is one templateIOC for each supported platform. A script, called "cpTemplate.bash", is also available which copies a specified template directory to create a new IOC with the specified name for the specified platform. The other script inputs perform some of the configuration tasks for the new IOC.

Fermilab is adopting the philosophy that macro substitution be performed at build time. Our motivation for this choice is to ensure that PVs be self documenting and forward link logic be available and visible for debugging.

For developers, we recommend, that after copying the template IOC, building it, and configuring it appropriately, one runs the IOC to test basic functionality and connectivity. The intent is to give new developers confidence from running a fully functional IOC prior to customizing it for the ultimate purpose of the IOC.

## CI/CD PIPELINE

The implementation of this model is automated in a CI/CD pipeline. Code management is performed with Github, as is documentation, and issue tracking. The Github Actions tool is invoked to automatically build and test the software, see Fig. 3, for a summary of the algorithm CI/CD algorithm.

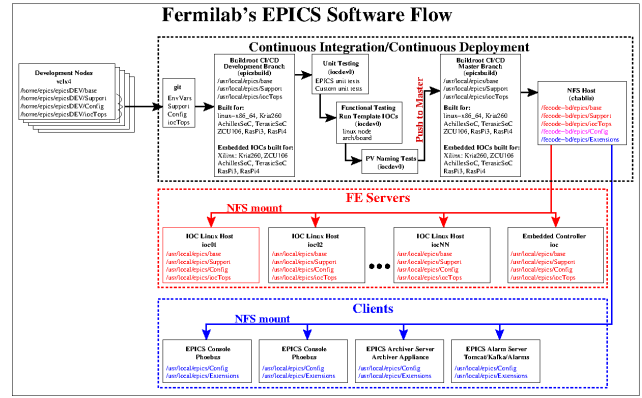

Figure 3: Software path from development to deployment on NFS host.

The pipeline follows a path triggered by a pull request from the developer on her/his account on the Controls network. Once the build is complete for each platform, a series of tests are performed: unit tests, functional tests on the template IOCs, and naming tests. If the tests are successful, a full build is performed and versioned and the the code is deployed to the NSF host. At this point, an IOC developer will have available on the Controls network, the NFS mount which hosts the directory:

/usr/local/epics/

from which all of the boiler plate epics code is available for each platform. The code version is preserved in the names of the directories, and so in the GitHub repositories. The softlinks "base" and "Support" are used to point to the latest versions of base-XXX and Support-YYY, so that user code need not modify build scripts when versions change. Three versions of the code are kept on the NFS host, so that only softlink change allows for a fall back to a previous version should problems with the build be observed.

The result of the build is that users now have fully functioning and tested boilerplate code available from an NFS mount for any of the supported platforms. In short, EPICS base is available as:

```
./base/bin/linux-x86_64
./base/bin/linux-arm_raspberrypi3
./base/bin/linux-arm_raspberrypi4
./base/bin/linux-arm_terasicsoc
./base/bin/linux-arm_achilles
```

```
./base/bin/linux-arm_zcu106
./base/bin/linux-arm_kr260
```

and

```
./base/lib/linux-x86_64
./base/lib/linux-arm_raspberrypi3
./base/lib/linux-arm_raspberrypi4
./base/lib/linux-arm_terasicsoc
./base/lib/linux-arm_achilles
./base/lib/linux-arm_zcu106
./base/lib/linux-arm_kr260
```

Similarly, all of the Support or Modules are available from the NSF mount. As an example of the asyn Support module, one has:

```
./Support/asyn/linux-x86_64
./Support/asyn/linux-arm_raspberrypi3
./Support/asyn/linux-arm_raspberrypi4
./Support/asyn/linux-arm_terasicsoc
./Support/asyn/linux-arm_achilles
./Support/asyn/linux-arm_zcu106
./Support/asyn/linux-arm_kr260
```

The IOCs are also processed through the same CI/CD pipeline. The script which copies a template IOC, configures the new IOC to be built for the selected platform. The output of the pipeline will install the new application in new subdirectory in:

`/usr/local/epics/iocTops`

Thus, all IOCs can be deployed from the production area on the NFS mount.

## TESTING IN THE CI/CD PIPELINE

To date, testing is the least developed part of our CI/CD pipeline. The testing application is a combination of Ruby and Pickle. There are 3 levels of testing in the CI/CD pipeline:

1. Unit tests – base, and many Support modules already come with unit tests. Here, our work will be to implement the testing for each platform.

2. Functional tests – the tests are performed at the templateIOC level.

3. PV name tests – tests for uniqueness of PV names and for ensuring that PV names follow our naming conventions.

After the first build of the code in the CI/CD pipeline, base, Support, and iocTops are deployed on a special server which is on the testing VLAN; this mimics the functionality of the production code NFS server. Also on the testing VLAN are the individual platforms which run their respective templateIOCs and go through the testing procedures. Successful testing triggers a production version build, see Fig. 3, which is deployed on production servers.

## ADDITIONAL TESTING

There are a variety of other tests that we will run which are not a part of our CI/CD pipeline. These include network stress tests and benchmarking bandwidth usage for different configurations.

There are also efforts to use Redis [8] as a cache of data from edge data acquisition systems upstream of EPICS; write and read access to Redis is being standardize so that we can create an EPICS module to populate normative types in a well defined and efficient way. When ready, this process will require testing for efficiency and reliability.

Another effort in progress is to run the IOC on the edge on the arm processor of a SoM using Userspace I/O [9] space to populate normative types directly from the FPGA memory. This, too, will require extensive testing for efficiency and reliability and perhaps benchmarking it against a Redis solution.

## DEPLOYING IOCS

We have 2 classes of IOC deployment:

### Soft IOCs

Since all of the software is built with the same dependencies and EPICS code base, all of the "soft" – non-embedded – IOCs will be deployed from Alma Linux servers. These IOCs will be launched using procServ on separate TCP ports and the procServ scripts will be launched from systemd. The log files for each IOC will be written to separate directories in:

`/scratch/epicsLogs/.`

At present, since all of the infrastructure and code base for each IOC is identical, we do not see added value in using containers for deploying these IOCs. Because each IOC is built only with the selected modules, there is no concern that the application is bloated with unnecessary code.

### Embedded IOCs

Our present list of supported embedded platforms includes Raspberry Pis and SoMs, namely: Achilles SoM, Terasic SoM, Xilinx ZCU106, and Xilinx Kria 260. The Raspberry Pi will be single process server, and so we treat it as an embedded system. For these platforms, an SD card or on board flash will have host specific network configuration and kernels, root file system, dtb files, etc. are loaded at boot time from a tftp server. We configure and use u-boot for network booting the embedded platforms. Like the soft IOCs, these boards will have the common NFS mount:

`/usr/local/epics.`

Also, like the soft IOCs, systemd will launch the IOCs via procServ and the log files will be written in separate directories in NFS mounted

`/scratch/epicsLogs/.`

*Older Hardware*

In the future, when present hardware becomes "old" hardware, we will likely implement containers to preserve older versions of software so as to preserve the older IOCs, until newer hardware becomes available.

## EPICS SERVICES

In addition to the infrastructure for IOCs, Fermilab is evaluating several EPICS services. Those presently under evaluation are:

- Operator HMIs – Phoebus
- Archiver Appliance
- Alarms
- Save & Restore
- Channel Finder

For these services, we will deploy them using a similar GitHub → CI/CD pipeline as is done for the EPICS base, Support, and IOC code. What goes into the GitHub repository are the configuration files for these services, and a push will trigger a CI/CD pipeline to perform sanity checks on the files before deploying them to the NFS server, in directories under:

`/usr/local/epics/Config`.

*Phoebus*

We have already accomplished significant evaluation of Phoebus and find it a very powerful and convenient tool for generating HMIs and using them for controls and monitoring. From the author's perspective, the real beauty of Phoebus is the hooks it provides for many different services such that operators have a single interface from which to launch their applications. The HMI files generated are referred to as "bob" files and are simply flat xml files. Though not in place yet, the CI/CD pipeline will include tests for these files to check that they meet Fermilab style requirements. Examples of some HMIs are shown in Fig. 4.

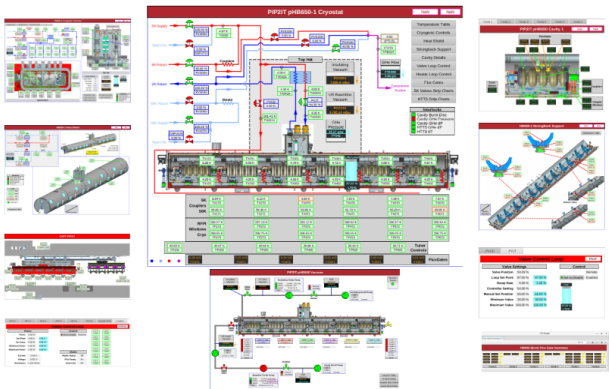

Figure 4: Several of the HMIs used for CM testing at PIP2IT.

From our experience, Phoebus is not without its limitations. As a java application, it demands significant resources and running it from a remote node requires great care. In our case, we are running it using Xpra [10] to run the X-clients on the remote host.

It is likely that Fermilab will also support other applications to run control and monitoring screens. As an example, we are evaluating DBWR (Display Builder Web Runtime) to allow the PIP2IT team to view read-only HMIs as viewed in the CMTF control room. Other applications being considered fall under the auspices of ACORN.

*Archiver Appliance*

The Archiver Appliance has also had significant evaluation to date, though we've not run more than a single appliance so far. We find the 3-tiered (short term, medium term, and long term) nature of archiving appealing. We also find its integration with Phoebus very useful and simple to implement. To date, we've not tested its capability to handle higher bandwidth data which may come from beamline instrumentation, for example.

We are actively using the Archive Appliance at PIP2IT, and it is likely that we will continue to use it in the PIP-II era. However, we continue to research other archiving options to suite our high bandwidth data archiving needs.

## TESTING AT PIP2IT

PIP2IT is the PIP-II Integration Teststand at the Cryomodule Test Facility (CMTF). This facility will be used to test PIP-II CMs prior to installing them in the superconducting LINAC.

From 2021 to June 2023, much of the control system infrastructure has been established for PIP2IT. EPICS IOCs were developed and several services were installed, including Phoebus for the HMIs, Archiver Appliance, Alarm Handling, Channel Finder, and Save and Restore.

EPICS IOCs were used in CM control and monitoring of the vacuum, CM cryogenics, HPRF, and some safety systems. They depended on the full infrastructure of the CI/CD pipeline and thus gave us ample opportunity to vet both the software and much of the CI/CD pipeline.

Several of the operator HMIs which are used at PIP2IT are shown in Fig. 4.

*Lessons Learned*

PIP2IT will run for the next few years, thus giving us ample opportunity to vet the software and services. During the first iteration, we had numerous iterations of the CI/CD pipeline scripts, and this was invaluable in teaching us to build a robust system.

In the recent CM tests, we did learn significant lessons regarding the services. The first is that services do not run reliably right out of the box. Each service requires one, or more, personnel to "own" the system. We had early failures with Phoebus and the Archiver Appliance, but we did manage to get them operating stably for most of the testing. Alarms, Channel Finder, and Save & Restore we never brought up to a reliable level. We intend to remedy this for the next CM tests in early 2024. The goal is that these systems be vetted and made robust, as well as to train operators, prior to establishing a reliable control system at PIP-II.

## OPERATION AT PIP-II

Figure 5 shows an early prototype the main operator display which may be used for PIP-II operations. The purpose of the page is to be both an accelerator status at a glance and a launcher for each complex system. For each CM, the status of each subsystem is displayed in stop-light style below each CM. Similarly, for beamline accelerator components and beamline instrumentation, a stop-light status is also displayed above the image of the PIP-II LINAC.

## SUMMARY

The PIP-II project will usher in a new era for Fermilab's high energy physics programs by providing a high power proton beam for a variety of experiments at Fermilab and to provide high flux neutrinos to experiments both at Fermilab and at long baseline neutrino experiments, such as DUNE.

The use of EPICS in the Fermilab's control system is now well established and will remain an integral part of it. In treating EPICS deployment as a green field and leveraging on using EPICS only for newer hardware allows for a single build of EPICS base and Support/Modules against which all IOCs are compiled. Thus the code is built and tested in a CI/CD pipeline before deployment to an NFS server from which the production IOC servers start the IOCs.

Several of the EPICS applications are presently under evaluation at Fermilab. It is likely that these services run in parallel with existing Fermilab applications, since EPICS will have to interface with the existing, native control system for the foreseeable future.

PIP2IT is the integration test stand where the PIP-II CMs will be tested before installing in the new superconducting LINAC. In addition to testing CMs, we are using the facility to develop and test EPICS controls and the services which are under consideration. We have already vetted many of the IOCs used to control the CMs in CM testing to date, as well as gained significant incite as to how we are to use several of the EPICS applications. We still have much more to learn, as we resume CM testing in early 2024.

## ACKNOWLEDGEMENTS

This manuscript has been authored by Fermi Research Alliance, LLC under Contract No. DE-AC02-07CH11359 with the U.S. Department of Energy, Office of Science, Office of High Energy Physics.

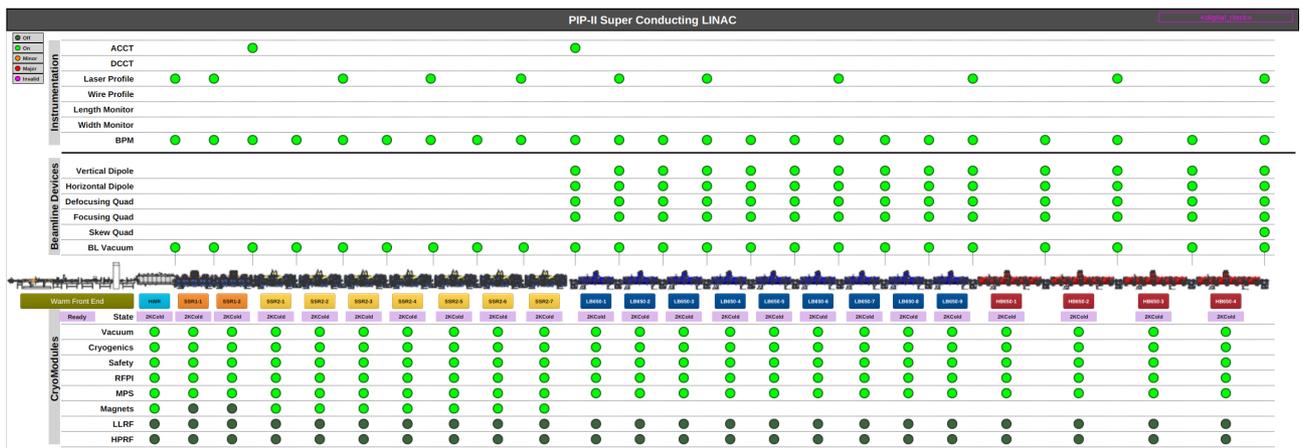

Figure 5: Prototype (simulation) operator main page for PIP-II accelerator.